\documentclass[prl,letterpaper,twocolumn,showpacs]{revtex4}
\usepackage{times,xspace}
\usepackage{amsbsy,amssymb,amsmath,bm}
\usepackage{graphicx,color,epsfig,rotate}
\usepackage{fancyhdr}

\def\bbbc{{\mathchoice {\setbox0=\hbox{$\displaystyle\rm C$}\hbox{\hbox
to0pt{\kern0.4\wd0\vrule height0.9\ht0\hss}\box0}}
{\setbox0=\hbox{$\textstyle\rm C$}\hbox{\hbox
to0pt{\kern0.4\wd0\vrule height0.9\ht0\hss}\box0}}
{\setbox0=\hbox{$\scriptstyle\rm C$}\hbox{\hbox
to0pt{\kern0.4\wd0\vrule height0.9\ht0\hss}\box0}}
{\setbox0=\hbox{$\scriptscriptstyle\rm C$}\hbox{\hbox
to0pt{\kern0.4\wd0\vrule height0.9\ht0\hss}\box0}}}}

\newcommand{\ignore}[1]{}
\newcommand{\mComment}[1]{}
\newcommand{\gComment}[1]{}
\newcommand{\jComment}[1]{}
\newcommand{\rComment}[1]{}
\newcommand{\lComment}[1]{}

\renewcommand{\mComment}[1]{\textcolor{blue}{Manny: #1}}
\renewcommand{\gComment}[1]{\textcolor{red}{Gerardo: #1}}
\renewcommand{\jComment}[1]{\textcolor{green}{Jim: #1}}
\renewcommand{\rComment}[1]{\textcolor{magenta}{Ray: #1}}
\renewcommand{\lComment}[1]{\textcolor{purple}{Rolando: #1}}

\begin{document}
\title{Field Induced Supersolid Phase in Spin-One Heisenberg Models}
\author{P. Sengupta$^{1,2}$ and C. D. Batista$^1$}
\affiliation{$^1$Theoretical Division, Los Alamos National Laboratory, Los Alamos, NM 87545 \\
$^2$ MST-NHMFL, Los Alamos National Laboratory, Los Alamos, NM 87545}

\date{\today}

\begin{abstract}
We use quantum Monte Carlo methods to demonstrate that the quantum phase diagram of the S=1 
Heisenberg model with uniaxial anisotropy contains an extended supersolid phase. 
We also show that this Hamiltonian is a particular case of a more general and ubiquitous 
model that describes the low energy spectrum of a class of {\it isotropic} and 
{\it frustrated} spin systems. This crucial result provides the required guidance for finding 
experimental realizations of a spin supersolid state.
\end{abstract}


\maketitle %
\thispagestyle{fancy}

Theoretical proposals \cite{Affleck90} for studying the Bose-Einstein condensation 
(BEC) with magnetic systems were followed by a vast number of experimental works 
\cite{Dimers}. These studies were done on spin dimer compounds for which the relevant 
$U(1)$ symmetry is ``protected'' by the intra-dimer singlet-triplet spin gap 
\cite{Suchitra06}. Magnetic systems have the advantage that the magnetic field, which  
plays role of the chemical potential, can be varied continuously over a large range 
of values. A natural question that arises is whether other phases that have been proposed for 
bosonic gases of atoms can be realized in quantum magnets. The supersolid (SS) state is 
a prominent and interesting example because the experimental evidence for this 
novel phase is still inconclusive \cite{Chan04}.

The search for the SS phase has motivated the study of different models for hard--core
bosons on frustrated lattices \cite{Triangular}. These models are  
relevant for gases of atoms in a periodic potential (substrate or optical lattice).
However, the spin $S=1/2$ Hamiltonians that are obtained from these models by applying a 
Matsubara--Matsuda transformation \cite{Matsubara56} are not relevant for real magnetic systems.
What makes these models unrealistic for magnetic systems is the large uniaxial exchange 
anisotropy. Moreover, the longitudinal and the transverse components of the exchange 
interaction have opposite signs: while the Ising interaction is antiferromagnetic (AFM),
the transverse exchange coupling is ferromagnetic. It is then natural and relevant to ask if a 
SS spin phase can exist in a magnetic system with isotropic (Heisenberg) interactions.
In this letter, we provide an affirmative answer to this question by calculating the 
quantum phase diagram of an $S=1$ spin--dimer Heisenberg model. The spin SS phase is induced by 
the application of a magnetic field whose Zeeman splitting is comparable to the magnitude 
of the exchange interactions.

To understand the physical origin of the spin SS, we shall start
by considering the simplest (although non-realistic) $S=1$ Hamiltonian that contains this phase in its 
phase diagram. This is an $S=1$--Heisenberg model with uniaxial single--ion and exchange anisotropies
on a square lattice:
\begin{equation}
H_H = J \sum_{\langle {\bf i,j} \rangle } (S^x_{\bf i} S^x_{\bf j} + S^y_{\bf i} S^y_{\bf j} 
+\Delta S^z_{\bf i} S^z_{\bf j} ) + \! \sum_{\bf i} (D {S^z_{\bf i}}^2 - B S^z_{\bf i})
\label{eq:H}
\end{equation}
where $\langle {\bf i,j} \rangle$ indicates that ${\bf i}$ and ${\bf j}$ are nearest 
neighbor sites, $D$ is the amplitude of the single ion-anisotropy and $\Delta$ 
determines the magnitude of the exchange uniaxial anisotropy. 
Note that although the exchange interaction is anisotropic, the longitudinal ($J$) and 
transverse ($\Delta$) couplings are both AFM (positive).
Henceforth, $J$ is set to unity and all the parameters are expressed in units of $J$.

\begin{figure}[]
\includegraphics[angle=0,width=8.5cm]{fig1.eps}
\includegraphics[angle=90,width=8.5cm]{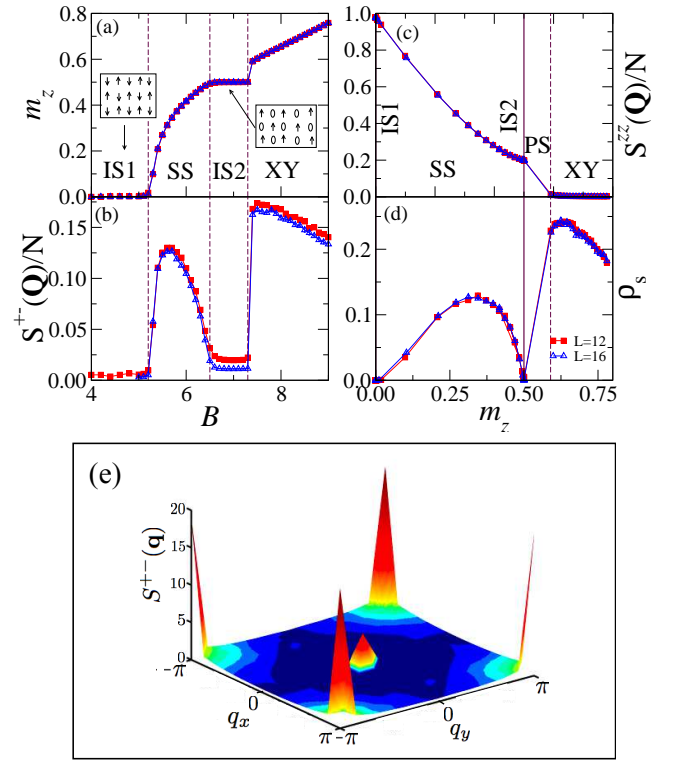}
\vspace{-1.8cm}
\caption{(Color online) Quantum phase diagram of $H_H$  
(Eq.~\ref{eq:H}) for $D=1.5$ and  $\Delta=1.8$. (a) Magnetization
as a function of field $B$. The SS phase appears between the 
two Ising (or solid) orderings denoted by IS1 and IS2. At higher fields,
there is a first order transition between the IS2 and the pure XY--AFM phases. 
(b) Square of the XY--AFM order parameter as a function of $B$.(c) and (d) 
Longitudinal component of the staggered SSF and stiffness as a function of the magnetization.
In a grand canonical ensemble, no ground state with $0.5 < m_z \lesssim 0.59$ (marked PS) 
is realized -- this corresponds to the discontinuous IS2-XY transition. For a 
canonical ensemble with magnetization in this range, the ground state phase
separates into spatial domains with $m_z=0.5$ and $m_z \approx 0.59$. (e)
Full momentum distribution of the form factor, $S^{+-}({\bf q})$. The peak
at ${\bf q}={\bf Q}$, in addition the one at ${\bf q}={\bf 0}$ indicate
that the off-diagonal order is modulated by the presence of simultaneous long-range
diagonal order.}
\label{fig:j18}
\end{figure}

The quantum phase diagrams for the spin models considered in this paper were obtained 
by using the Stochastic Series expansion (SSE) quantum Monte Carlo (QMC) method. The 
simulations were carried out on a square lattice of size $N=L\times L$, 
with $ 8 \leq L \leq 16$. 
We find rapid convergence with $N$ for the system sizes 
studied (see Fig.~\ref{fig:j18}). Since the SSE is formulated in the grand 
canonical ensemble, the simulations are performed at fixed magnetic field,
instead of fixed magnetization.

As the external field, $B$, is varied, the ground state of $H_H$ goes through a number
of phases, including spin-gapped (IS) phases with Ising-like ordering and 
gapless (XY) phases with dominant XY-ordering. The IS phases are characterized
by long-range (staggered) diagonal order measured by the longitudinal 
component of the static structure factor (SSF),
\begin{equation}
 S^{zz}({\bf q})={1\over N}\sum_{j,k}e^{-i{\bf q}\cdot({\bf r}_j-{\bf r}_k)}\\ 
\langle  S^z_jS^z_k\rangle. 
\end{equation}
The XY--phase has long-range off-diagonal ordering measured by the 
transverse component of the SSF,
\begin{equation}
 S^{+-}({\bf q})={1\over N}\sum_{j,k}e^{-i{\bf q}\cdot({\bf r}_j-{\bf r}_k)}
\langle S^+_jS^-_k \rangle.
\end{equation}
The XY--ordering is equivalent to a Bose--Einstein condensation (BEC) whose 
condensate fraction is equal to $S^{+-}({\bf Q})$ with ${\bf Q}$ 
being the ordering wave--vector (${\bf Q}=(\pi,\pi)$ for the case under consideration). 
The superfluid density corresponds to the 
spin stiffness, $\rho_s$, defined as the response of the system to a twist in the boundary
conditions. The stiffness is obtained from 
the winding numbers of the world lines ($W_x$ and $W_y$) in the $x$- and
$y$- directions: $\rho_s=\langle W_x^2+W_y^2\rangle/2\beta$. 

The IS (XY) phase is marked by a diverging value of $S^{zz}({\bf Q}) \propto N $  
($S^{+-}({\bf Q}) \propto N $) in the thermodynamic limit $N \to \infty$. In addition,
$\rho_s$ vanishes in the gapped IS phase while it is finite in the gapless XY phase.
A spin SS phase is characterized by a finite value of both
$S^{zz}({\bf Q})/N $ {\em and\/} $\rho_s$. Both
quantities are always finite for finite size systems and estimates for $N \to \infty$
are obtained from finite-size scaling.

Fig.~\ref{fig:j18} shows the quantum phase diagram as a function of  
magnetic field, $B$, for $D=1.5$ and $\Delta=1.8$. For clarity 
of presentation, $S^{zz}({\bf Q})$ and $\rho_s$ are plotted as a function
of the resulting magnetization $m_z$. The $m_z (B)$ curve features two prominent
plateaus corresponding to different IS phases. For small $B$, the ground state is a gapped
AFM solid (IS1)  with no net magnetization. The stiffness, $\rho_s$,
and  $S^{+-}({\bf Q})$ vanish in the thermodynamic limit, while $S^{zz}({\bf Q})/N$ 
is slightly smaller than $1$ because the spins are mainly in the  
$S^{z}_{{\bf i}}=\pm 1$ states depending on which sublattice they belong
to. The magnetization stays zero up to the critical field, $B_{c1}$,
that marks a second order BEC
quantum phase transition (QPT) to a state with a finite fraction of spins 
in the $S^z_{{\bf i}}=0$ state. This state has a finite $S^{zz}({\bf Q})/N$ as well 
as finite $\rho_s$ and $S^{+-}({\bf Q})/N$, i.e., SS order. The diagonal order results
from the $S^z_{{\bf i}}=\pm 1$ 
sublattices while the off-diagonal order arises out of a BEC of the 
flipped spins ($S^z=0$ ``particles''). The magnetization increases continuously up
to $B\approx 6.4$, where there is a second BEC--QPT to a   
second Ising-like state (IS2) where all the $S^z_{{\bf i}}=-1$ have been flipped to the 
$S^z_{{\bf i}}=0$ state.  $S^{zz}({\bf Q})/N$ remains divergent for $N
\to \infty$, but
the stiffness, $\rho_s$, and condensate fraction, $S^{+-}({\bf Q})$, drop to zero. The ground 
state remains in the IS2 phase for $6.4 \lesssim B \lesssim 7.2$. Upon further
increasing the field, there is a first order transition to a pure XY--AFM 
phase ( $m_z$ changes discontinuously from $m_z=0.5$ to $m_z\approx 0.59$).
In the grand canonical ensemble, no ground state with any intermediate value
of the magnetization is realized. For a canonical ensemble with a fixed magnetization
$-0.6 < m_z < -0.5$, the ground state will phase separate into IS2 and XY regions with $m_z=0.5$
and $m_z=0.59$. In the pure XY phase, the diagonal order vanishes while
$\rho_s$ and  $S^{+-}({\bf Q})/N$ remain finite. This situation persists
until all the spins have flipped to the $S^z_i=1$ (fully polarized) state. 

Further insight into the SS phase is obtained from the
momentum dependence of $S^{+-}({\bf q})$  ( Fig.~\ref{fig:j18}(e)).
The peaks at ${\bf q}=(0,0)$ and ${\bf q}={\bf Q}$
indicate that the off-diagonal long range order is modulated by the presence of 
solid order. This confirms that the SF component of the SS
phase results from a BEC of $S^z_{{\bf i}}=0$ spin states 
that occupy the $S^z=-1$ or down--sublattice with higher probability. This 
feature distinguishes the SS phase from a uniform canted AFM phase. In 
experiments with magnetic materials, the components of the structure factor
are selectively measured using standard polarized neutron scattering experiments.

\begin{figure}[]
\includegraphics[angle=0,width=8.5cm]{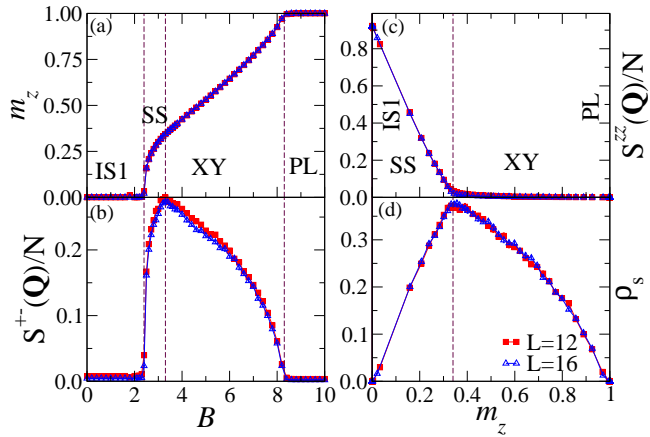}
\vspace{-0.3cm}
\caption{(Color online) Same as Fig.~\ref{fig:j18}, but with parameters $D=1.5, \Delta=1.2$. The
second magnetization plateau disappears completely. Instead, there is a direct
(continuous) SS-XY transition. At high fields, there is a transition to a fully
polarized state (PL).}
\label{fig:j12}
\end{figure}

For smaller values of $\Delta (< D)$, the second magnetization plateau disappears
completely (Fig.\ref{fig:j12}) leaving a second order transition 
from the SS to the XY--phase. Consequently, there is no phase separation regime. 
The extent of the SS phase decreases with decreasing 
$\Delta$ and vanishes for $\Delta \approx 1$.

We shall now discuss the relevance of these results for finding a SS phase in 
real magnets. We note that although a U(1) 
invariant model provides a good description of spin compounds whose 
anisotropy terms are very small compared to the Heisenberg interactions, this 
invariance is never perfect.  The transition metal magnetic 
ions belong to this class because the spin-orbit interaction is much smaller than 
the crystal field splitting. These spin systems have small 
exchange anisotropies for the same reason. Therefore, models that assume opposite 
signs for $J_z$ and $J_{\perp}$ \cite{Triangular} or large values values of 
$J_{\perp}/J_z$ \cite{Ng06} are not applicable to these spin compounds. 
We will show below that it is not necessary to assume a strong uniaxial exchange 
anisotropy for obtaining a SS phase.

\begin{figure}[!htb]
\hspace*{-1.1cm}
\hspace*{1cm}
\includegraphics[angle=90,width=9cm]{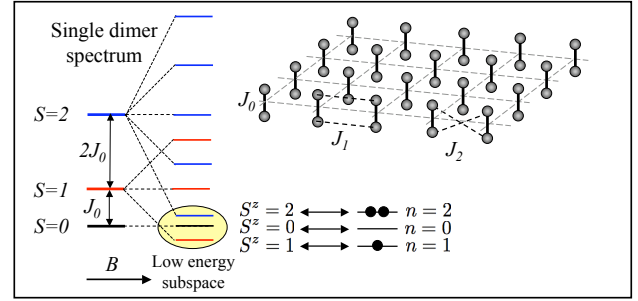}
\vspace{-2.8cm}
\caption{(Color online) Square lattice of S=1 dimers with an intra-dimer Heisenberg
AFM interaction $J_0$ and inter-dimer interactions $J_1$ and $J_2$. The left 
side shows the low energy subspace of the single dimer spectrum in the presence of a
magnetic field.}
\label{dimer}
\end{figure}

The system to be considered is a square lattice of $S=1$ dimers
(Fig.~\ref{dimer}) which only includes {\it isotropic} (Heisenberg) AFM
interactions, an intra-dimer exchange $J_0$ and inter-dimer {\it frustrated} couplings $J_1$ and $J_2$: 
\begin{eqnarray}
H_D &=& J _0\sum_{\bf i} {\bf S}_{\bf i+} \cdot {\bf S}_{\bf i-} +
J_1 \sum_{\langle {\bf i,j} \rangle, \alpha}  {\bf S}_{\bf i\alpha} \cdot {\bf S}_{\bf j\alpha}
\nonumber \\
&+& J_2 \sum_{\langle {\bf i,j} \rangle, \alpha}{\bf S}_{\bf i\alpha} \cdot {\bf S}_{\bf j{\bar \alpha}} 
- B \sum_{{\bf i}\alpha} S^z_{\bf i \alpha}.
\end{eqnarray}
The index $\alpha=\pm$ denotes the two spins on each dimer. The single dimer spectrum consists
of a singlet, a triplet and a quintuplet (see Fig.~\ref{dimer}). The energy difference between the 
singlet and the triplet is $J_0$, while the difference between the quintuplet and the triplet is $2J_0$.

For $J_1,J_2 \ll J_0$, the low energy subspace of $H_D$ consists of the singlet,
the $S^z=1$ triplet and the $S^z=2$ quintuplet (see Fig.~\ref{dimer}).
The low energy  effective model, $H$, that results from restricting $H_D$ to this subspace is 
conveniently expressed in terms of {\it semi--hard--core} bosonic operators, $g^{\dagger}_{\bf i}$ and 
$g^{\;}_{\bf i}$, that satisfy the exclusion condition ${g^{\dagger}_{\bf i}}^3=0$ 
(no more than two per site) \cite{Batista00,Batista04} and
obey the commutation relations of canonical bosons except for the 
commutator $[ g^{\;}_{\bf i} , g^{\dagger}_{\bf j} ] = \delta_{\bf i,j} (1-n_{\bf i})$
($n_{\bf i}=g^{\dagger}_{\bf i}g^{\;}_{\bf i}$ is the number operator). The expression of $H$ in 
terms of these operators is:
\begin{eqnarray}
H &=& \frac{1}{2} \sum_{\langle {\bf i,j} \rangle} 
(g^{\dagger}_{\bf i}g^{\;}_{\bf j}+g^{\dagger}_{\bf j}g^{\;}_{\bf i} ) (h_1+h_2+h_3)
- \mu \sum_{\bf i} n_{\bf i} 
\nonumber \\
&+& \frac{U}{2} \sum_{\bf i} n_{\bf i} (n_{\bf i}-1) + V \sum_{\langle {\bf i,j} \rangle} (n_{\bf i}-1) (n_{\bf j}-1)
\label{eq:Heff}
\end{eqnarray}
with $h_1=t_1 (n_{\bf ij}-2)(n_{\bf ij}-3)$, $h_2=2t_2 (n_{\bf ij}-1)(3-n_{\bf ij})$,
$h_3=t_3 (n_{\bf ij}-1)(n_{\bf ij}-2)$, and $n_{\bf ij}=n_{\bf i} + n_{\bf j}$. 
The amplitudes $t_1$, $t_2$ and $t_3$ correspond to single-particle hopping terms when 
there are one, two or three particles respectively on the corresponding bond $\langle 
{\bf i,j} \rangle$. The case $t_1=t_2=t_3=t$ corresponds to the  bosonic Hubbard model 
with n.n. repulsion \cite{Sengupta05} in a truncated Hilbert space. 
Our $S=1$ Heisenberg Hamiltonian with uniaxial anisotropy, $H_H$, 
is obtained for $U=D$, $V=\Delta J$, $\mu=D+B$ and $t_j=\sqrt{2}J/2^{j/2}$ with $j=1,2,3$ after
we map on each site the eigenstates of $S^z_{\bf i}$ onto the eigenstates of 
$n_{\bf i}$: $S^z_{\bf i}=n_{\bf i}-1$ and $S_{{\bf i}}^+ = g_{{\bf i}}^{\dagger} [\sqrt{2} + (1-\sqrt{2})n_{\bf i}]$.

\begin{figure}[]
\includegraphics[angle=0,width=8.5cm]{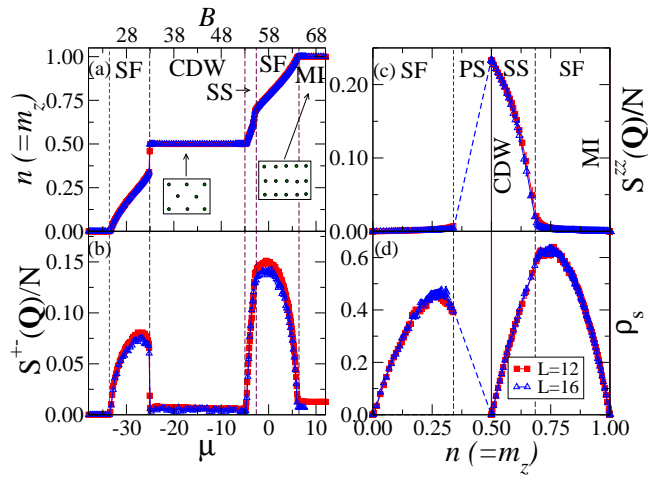}
\vspace{-0.3cm}
\caption{(Color online) Quantum phase diagram of $H$(\ref{eq:Heff})
for  $U=30.0, V=7.0$. 
(a) Particle density $n$ or $m_z$ as a function of the 
chemical potential $\mu$ (lower axis) or field $B$ (upper axis). 
(b) Condensate fraction or square of the AFM--XY order parameter. (c) and (d) The staggered SSF 
and stiffness as a function of $n=m_z$. The range of densities marked
PS is inaccessible in the grand canonical ensemble and would result in a phase separated
state in a canonical ensemble.}
\label{fig:ebhm}
\end{figure}

As we mentioned before, $H$ also describes the low energy spectrum of $H_D$. In this 
case, we have $U=J_0$, $V=(J_1+J_2)/2$, $\mu=B-J_0-z(J_1+J_2)/2$ and 
$t_j=8 a^j(J_1-J_2)/3\sqrt{3}$ with $a=\sqrt{3}/2$,
after mapping the eigenstates of $S^z_{{\bf i}+} + S^z_{{\bf i}-}$ into the eigenstates of $n_{\bf i}$ 
by the simple relations: $n_{\bf i}=S^z_{{\bf i}+} + S^z_{{\bf i}-}$ (see Fig.\ref{dimer}) and 
$g_{{\bf i}}^{\dagger} = (S^+_{{\bf i}+} - S^+_{{\bf i}-})[{\frac{\sqrt{3}}{2\sqrt{2}}} + (1-{\frac{\sqrt{3}}{2\sqrt{2}}})S^z_{\bf i}]$. 
Fig.~\ref{fig:ebhm} shows the quantum phase diagram as a function of $\mu$ (or $B$) for
$U=30.0$ and $V=7.0$ (in this case we take $(J_1-J_2)/2$ as the unit of energy). This set of parameters corresponds to  
$J_0=30$ $J_1=8$ and $J_2=6$ that satisfies the conditions $J_0 > z(J_1+J_2)/2$ and
$J_0 \gg z(J_1-J_2)/2$  necessary for the validity of $H$ as a low energy effective model for $H_D$. 

At small $\mu$ or $B$, the empty state (all the dimers in a singlet state) has the lowest energy.
For $\mu > \mu_{c1}$ ($B>B_{c1}$) a finite density of bosons (triplets) is stabilized in the ground state
giving rise to a BEC (XY--AFM ordering) at $T=0$ with a finite the stiffness $\rho_s$.
The absence of solid (Ising) ordering is indicated by
$S^{zz}({\bf Q})/N \rightarrow 0$. The density (magnetization) increases monotonically as a function
of $\mu$ or $B$ until $\mu=\mu_{c2} \approx 2.9$ where there is a discontinuous transition to a charge-density wave (CDW)
or Ising-like phase with $n=m_z=0.5$ (the dimers of one sublattice are in a 
triplet state while the other dimers remain in the singlet state). For $\mu > \mu_{c3} \sim 23.4$, 
some of the dimers of the singlet sublattice are turned into triplets that propagate primarily
on the singlet sublattice ($U \gtrsim z V$ where $z=4$ is the coordination number). Consequently, there is
a BEC--QPT (broken U(1) symmetry under rotations around the $z$--axis) in ${\cal D}=d+2$ dimensions to a SS phase, where 
$d$ is the spatial dimensionality. 
The diagonal or solid order disappears at an Ising--like quantum critical point
in ${\cal D}=d+1$ dimensions for $\mu=\mu_{c4} \approx 25.4$  (broken $Z_2$ symmetry of
translation by one lattice parameter followed by a $\pi$--rotation around the $z$-axis). Upon further increase in $\mu$, 
the filling increases monotonically in the resulting SF phase till the ground state enters a Mott insulating (MI) phase with all the dimers  
in the triplet state. 

The mechanism for the formation of the SS phase is explained most readily in
the bosonic language \cite{Sengupta05}. In the strong coupling limit ($U,V \gg t$), the half-filled ground state
($n={1\over 2}$) is a checkerboard solid  (one sublattice is
single occupied while the other sublattice is empty). Doping away from 
$n={1\over 2}$ results in different scenarios depending on the nature of doping and
the relation between the coupling constants $U$ and $V$. {\it Extracting} bosons from the 
$n=1/2$ crystal costs chemical potential $\mu$ but no potential energy. The
kinetic energy gain of the resulting holes is quadratic in $t$ for isolated holes (${\cal O}(t^2/V)$), but becomes {\em  linear\/} 
in $t$ if the holes segregate in a SF bubble.  Consequently, if the total density is fixed, the system separates 
in a commensurate crystal with $n=1/2$ and and a uniform SF region with
$n<1/2$. This implies a first order transition between the solid and the SF phases as 
a function of $\mu$ (see Figs. \ref{fig:j18} and \ref{fig:ebhm}).

Doping of the $n=1/2$ crystal with {\it additional} bosons works
differently depending on the relation between $V$ and $U$.  The energy
cost to place a boson at an empty (occupied) site is $E_0\equiv
zV-\mu$ ($E_1\equiv U-\mu$).  Respectively, for $U\gg zV$, the additional
bosons fill empty sites and mask the checkerboard modulation; for
$U-zV\gg |t|$ the situation is precisely particle-hole conjugate to
hole doping. In particular, in the hard-core limit $U\to\infty$, the crystalline order
is always unstable for $n\neq1/2$. However, for $zV \sim U$, the bosons can be placed on either an
occupied or unoccupied site.  The kinetic energy gain of the added boson
is now linear in $t$ because the potential barrier, $|zV - U|$, for moving the 
bosons to nearest neighbors is not much bigger than $t$.
As a result, the added bosons form a SF phase on top of the density wave background and
hence the ground state has simultaneous solid and SF
orders.  This SS phase is stable for a sufficiently small concentration of 
added bosons. This is confirmed by the quantum phase diagram shown 
in Fig.\ref{fig:ebhm} where the SS phase appears right next to the $n=1/2$ CDW. 
We emphasize that this phase requires to have two bosons on the same site, which is not
possible for hard-core bosons (or, equivalently, for $S={1\over 2}$ spins).

In summary, we have shown that simple two--dimensional $S=1$ Heisenberg models have a spin SS 
ground state induced by magnetic field. The physical mechanism that leads to this phase does not depend
on the dimensionality and similar results are expected for three and one--dimensional lattices 
\cite{Batrouni06,Pinaki07}. In particular, by finding a SS phase in an isotropic S=1 Heisenberg model we are 
providing the required guidance for finding this novel phase in real spin systems. The crucial ingredients 
for the described mechanism are: $S=1$ spins, a {\it dimerized} structure and a {\it frustrated} inter--dimer 
coupling.


 

LANL is supported by US DOE under Contract No. W-7405-ENG-36.

\end{document}